\documentclass[graybox]{svmult}

\usepackage{type1cm}        
\usepackage{makeidx}         
\usepackage{graphicx}        
\usepackage[inkscapelatex=false]{svg}            
\usepackage{multicol}        
\usepackage[bottom]{footmisc}

\usepackage{newtxtext}       %
\usepackage[varvw]{newtxmath}       

\makeindex             

\begin{document}


\title*{Cost Control and Efficiency Optimization in Maintainability Implementation of Wireless Sensor Networks based on Serverless Computing}
\author{Tiannan Gao and Minxian Xu}
\institute{Tinanan Gao \at  Shenzhen Institute of Advanced Technology, \email{tn.gao@siat.ac.cn}
\and Minxian Xu (corresponding author)\at  Shenzhen Institute of Advanced Technology \email{mx.xu@siat.ac.cn}}
%
%
\maketitle

\abstract*{Each chapter should be preceded by an abstract (no more than 200 words) that summarizes the content. The abstract will appear \textit{online} at \url{www.SpringerLink.com} and be available with unrestricted access. This allows unregistered users to read the abstract as a teaser for the complete chapter.
Please use the 'starred' version of the \texttt{abstract} command for typesetting the text of the online abstracts (cf. source file of this chapter template \texttt{abstract}) and include them with the source files of your manuscript. Use the plain \texttt{abstract} command if the abstract is also to appear in the printed version of the book.}

\abstract{Wireless sensor network (WSN) has been developed for decades and have performed well in the performance, power consumption, and congestion control. However, the following problems have not been addressed, such as inaccurate cost estimation of device’s lifecycle, highly-coupled engineering development, and low utilization of hardware and software resources during the life cycle of WSN. Therefore, we first propose the conceptual view of maintainability implementation for WSN based on Serverless Computing. The maintainability implementation refers to the ability to meet the WSN product to consume the minimum resources with a higher probability in configuration, trial production, debugging, batch production, deployment, operation, and maintenance phases. And then, we discuss that Serverless Computing can be realized at the software functional level of WSN to decouple the device operation and functional development, greatly improve the reuse of resources and exclude the hardware interference.
 From the perspective of maintainability and cost control, the concept of Serverless Computing can be used to build WSN platforms, which can support the functions of data collection and data management into functional development that may benefit from exploration through upfront expenditures, thereby significantly reducing design, manufacturing, and operational costs.
Finally, based on existing technologies and smart city scenarios, the idea of a WSN platform for Serverless Computing is given with a case study.}

\section{Introduction}
\label{sec:1}
Wireless sensor network is a paradigm composed of a large number of stationary or mobile sensors in a self-organizing and multi-hop manner to collaboratively sense, collect, process and transmit information about sensed objects in the geographic area covered by the network and eventually send this information to the owner of the network.
Its emergence was based on the industrial need for large-scale, high-efficiency, low-cost control and acquisition and the continued technological advances in radio frequency (RF), processors, nanotechnology and microelectromechanical systems (MEMS) \cite{MEMS-acm}. 
It has been widely used in industrial monitoring, high-volume intelligent meter reading, parking lot space monitoring and other scenarios, which shows the maturity of WSN paradigm.

Recent research hotspots in this area are centered on media access control (MAC) layer communication protocols, functional boundaries, energy consumption control, security and forensics. The communication protocols of WSN have been well studied. M. Zimmerling   et al. \cite{sync-proto-survey-acm} have summarized more than 40 recently used synchronous MAC layer protocols, and the protocols were divided into nine categories according to one, many, all, and two-to-two correspondence (e.g., one-to-many, many-to-many, etc.), and the performance of the protocols in their categories was evaluated. Liu  et al. \cite{isac-limit} summarized the future 5G environment with multiple input and multiple output (MIMO) antenna conditions, Integrated Sensing and Communication (ISAC) system can be summarized as device-free sensing, device-based sensing, device-free ISAC and device-based ISAC, four kinds of architecture. It is also concluded that the fundamental limitation of the ISAC channel cannot be obtained by employing a simple combination of existing reach performance boundary techniques in sensing and communication systems alone. In terms of energy consumption control, Osamah Ibrahim Khalaf  et al. \cite{energy-consumption-control} summarized in the paper that some nodes have lower lifetime than other nodes. In the WSN structure of star topology, cluster head nodes, nodes near sink gateway, and multi-hop nodes with high frequency transmission have aggregated low lifetime due to higher operating frequency relative to other nodes, called hotspot issue. And then they summarize historical proposed solutions to mitigate the problem and provide a platform for researchers exploring solutions to these topical problems in the network and proposing novel solutions. In terms of Internet of Things (IoTs) security and forensics, Maria Stoyanova et al. \cite{iot-forensics-survey} summarize that at this stage, IoT devices such as WSN have evolved to the point where vendors need to provide compliant forensic system designs for their devices, both at the legal level and in terms of technical means. Therefore WSN design solutions that meet the needs of legal forensics may also be an opportunity and challenge for the future.

In engineering applications, WSN has been used in home automation, environmental monitoring, various control systems and other application scenarios. In terms of home automation scenarios, the market is very mature for self-organized communication among various smart appliances, wearable devices and other IoT devices, and mesh networking and smart interconnection among devices have been widely used. In terms of environmental detection, WSN is commonly used as a monitoring execution unit for forest fires, hydrological monitoring, and etc. Taking the production environment in the rolling mill as an example, the application scenario in which its control system is located has the following characteristics: construction workers face high safety risks, high costs of temporary stops during operation, and closely linked production links. Therefore, the sensor information transmission required by its control system should meet the safety production specifications of large-scale collaborative production lines, timely information transmission, and reliable transmission methods. With the low cost of WSN and low coupling with the production environment, stable control and continuous monitoring of the production line can be achieved.

WSNs have following common features. First, the WSN system design is highly targeted, and additional requirements need to be re-established for the next life cycle. Second, the cost of deployment and maintenance is high, therefore, in engineering practices, the developers often use redundant deployment instead of subsequent maintenance to reduce costs. These characteristics guide the implementation strategy often results in WSN material resource waste, channel interference and environmental pollution problems.

In the future direction of WSN, many future directions and challenges \cite{future-chlg} have been identified, such as soft computing (edge computing), device security, heterogeneous interoperability, self-organizing protocols, routing schemes for managing IoT networks, data management, etc. These issues are essentially trade-offs between WSN functionality and actual lifetime versus cost and resources. The idea proposed in this paper is to synthesize the above challenges and problems of WSN and to provide an approach to visualize the methods and frameworks into engineering implementation plans based on the concept of \textit{Serverless Computing} in the face of the maintainability assessment methods and decision frameworks that have been constructed in the WSN lifecycle, and then verify whether the expected benefits of maintainability are achieved through engineering practice or not.

In this paper, we illustrate the concept of maintainability in WSN and try to quantify and analyze the links of their lifecycle to achieve controllable expenditures in each stage of their lifecycle in order to improve the expected benefits. And then we use the concepts of Serverless Computing for user participation and control granularity segmentation to build an engineering implementation method to achieve WSN maintainability. The method is, based on the current stage of embedded devices per unit price of performance improvement, cost per unit area of power consumption decline, explore the resulting expected revenue space if it can be achieved with a greater probability, it allows to improve the design phase of expenditure on the use of the system's resources (control, acquisition, computing, WSN reconfigurable architecture, etc.) programmable design to achieve resource-based sale strategy division. The efficient use of WSN resources in the loop can reduce the coupling between the operation and optimize the maintenance phase and the production environment, amortize the cost with pre-taken increased revenue, and achieve true environmental protection through the efficient use of WSN system.

We propose the maintainability implementation of WSN based on the challenges and problems as above and study the strategies for cost resource allocation throughout their lifecycle. In this paper, we first introduce the definition, characteristics, and existing challenges of WSN maintainability implementation in Section 2 and the introduction of Serverless Computing in Section 3. In Section 4, by combining the existing hardware and software technologies, we illustrate the idea of increasing the equipment utilization rate to amortize the research and development (R\&D) cost based on Serverless Computing, and propose a feasible model for the whole process of maintainability under the existing technology framework. In addition, the full-flow application is demonstrated in a smart city scenario. Finally, the conclusion and future work are summarized in Section 5.

\section{WSN Maintainability Implementation}
\label{sec:2}
At this stage, the configuration, trial production, debugging, batch production, deployment, operation, and maintenance processes of WSN are still dominated by project requirements, and the coupling in the whole process is so high that the resource utilization in the overall life cycle of WSN is not high and the costs are difficult to accurately predict and control. In order to predict the cost in a certain confidence interval before project initiation, guide the decision making in project practice, reduce the coupling between process links, improve the reusability of resource spending, and provide experience for the next project, we propose the concept of maintainability in WSN. This section covers a practical engineering approach to maintainability called WSN maintainability implementation.

\subsection{Definition}
\label{subsec:2}

%
WSN maintainability implementation satisfies a prerequisite, which aims to meet a ten-year life expectation of the WSN product throughout WSN maintainability implementation phases conversion.
Its phases include configuration, trail production, debugging, batch production, deployment, operation, and maintenance. 
The phase conversion decision is based on a trade-off between the confidence in the net benefit and the confidence in the successful implementation of the program. 
The net benefit refers to the expected benefit of the corresponding solution in each phase subtracting the expected cost and resource consumption. 
Successful implementation means that the implementation of the phase is completed exactly as scheduled and the post-implementation metrics are met, e.g., the WSN product debugging low power current reaches 80\% percent of the chip datasheet within 7 days of the debugging phase.
Confidence refers to the probability of achieving the desired goal for both in each phase. 
We usually expect the cost and resource consumption in each phase to be as low as possible among all possible solutions, and the expected benefits to be as high as possible, and the implementation to be executed successfully, and the confidence level of the above factors to be as high as possible. In practice, however, the tendency of the confidence level of both is often chosen according to the actual needs. 
The cost refers to the money spent, the resources refer to the manpower, material resources, information channels that can be mobilized in the project process, and there is an intersection between the two. Finally, the resources that can be converted to cost participate in the statistics at the corresponding cost.

\subsection{Life cycle phase}

Based on the tasks and characteristics of each phase, we divided the WSN maintainability implementation life cycle into seven parts.

\subsubsection{Configuration}
Configuration refers to the analysis of its needs after the project is established, based on previous engineering experience or feedback information resources from other phases of this project, the expected assessment of costs and resources, taking into account the factors of the specific environment, weighing the cost, reliability, upgrade redundancy, technology selection (the cost of technical debt for later upgrades) expenditures, the results of other phases, and repeatedly design the hardware and software options for WSN products The estimated cost expenditure has a high confidence level. Reliability, upgrade redundancy, and technology selection are the factors that affect the cost. The design process should consider the balance between actual needs and costs, such as high circuit reliability requirements for applications, which should be made in the design phase of circuit redundancy design, up the material cost budget. Upgrade redundancy should consider the hardware needs in ten years can be, should not be too high. Technology selection should also consider the trade-off between reliability and upgrade redundancy, as well as the advantage of increased redundancy due to technology upgrades over a ten-year cycle versus the cost of additional expenditures.

\subsubsection{Trial production}

After completing the configuration phase, small-scale trial production in accordance with the design is used as the basis for entering the debug phase after the verification of good products. Trial production is often used to verify the gap between the actual effect of the circuit and the simulation results, and to troubleshoot initial failures, such as Cooja Simulator \cite{cooja} is the developing and debugging application of Contiki OS, which is an operating system for the Internet of Things. Essentially it's a smaller cost outlay to avoid as much as possible the larger expense of scaling up errors in the mass production phase. For example, small batches are made for each option while the product is tested in the lab for electrical stability.

\subsubsection{Debugging}

After the electrical performance of the product is verified, the circuit operating parameters are tuned for the datasheet indicators as a benchmark. The essence of the same is to get more guidance information in this project at a smaller cost and resource expenditure. For example, the center frequency offset of the radio frequency (RF) chip, the received signal strength indication (RSSI) can reach the dB indicator of the datasheet. Whether the low-power current of the control part can reach the minimum standard of the datasheet. If not, what is the maximum low-power current that can meet the runtime, and what is the corresponding cost overhead for each reduced current.

\subsubsection{Batch production}

After the first three phases are largely completed or the unexpected costs of the later phases are within certain expectations, trade-offs are made between alternative designs, mass production vendors, materials, yields, redundant production, and costs. Based on the trade-offs, batch production is performed. Yield testing and lab environment validation of all products are performed after batch production. In case of large scale problems go back to the configuration phase at a significant cost. For example, radio frequency (RF) modules are to be shielded in places with strong electromagnetic interference. For projects with high maintenance costs in the field, the circuit design should be a highly reliable and slightly more costly solution in exchange for a higher probability of controlling the cost of the maintenance phase at a lower level.

\subsubsection{Deployment}

After passing mass production verification, package the device to field deployment testing. The process takes full consideration of the actual environment, utilizing the redundant design from the previous phase and prioritizing trying low-cost solutions. For example, we prioritize measuring the communication environment before deployment and switch to the channel with the least interference. For the problem of communication failure in some special locations, priority is given to trying to move the location to exclude the possibility of a multi-path effect.

\subsubsection{Operation}

After deployment, it works properly, continuously monitors the operational data and analyzes the gap between the actual behavior of WSN packets and the expected behavior. There also needs a corresponding maintenance plan for possible failures. For example, Guo et al. \cite{sniffer} designed a listening method for a WSN protocol based on Dong et al. \cite{mac-protocol} implemented a packet behavior collection analysis based on the listening method.

\subsubsection{Maintenance}

Target problem and requirement improvements based on actual packet behavior versus expected gaps until the end of the 10-year life cycle. Add new requirements within hardware cost constraints. Ongoing lessons learned, configuration, debugging, and production phase work can be reused and reduced by subsequent projects, and deployment and operation phases to maximize equipment utilization.

\subsection{Resources and costs}

Resources and costs as sufficient conditions for driving links in the WSN life cycle. We give the scope of resources and costs as follows.

\subsubsection{Resources}

Resources are objectively existing substances that are directly used in the WSN life cycle. Resources include, but are not limited to, information, manpower, and supplier channels. There is a mutual influence relationship between them and costs. For example, information affects the proportion of cost allocated to resources, and cost affects supplier channels. Some of these resources can eventually be counted as cost.

\subsubsection{Cost} 

Cost refers to the time and money spent during the project cycle, and is used as an indicator to evaluate the final maintainability design. Cost control is achieved mainly through the ideas of increasing revenue and cost-cutting. Cost-cutting refers to reducing the budgeted cost of the design under the requirement of stability and reliability, while increasing revenue refers to improving the utilization of resources that have already been paid for. Serverless Computing is an implementation of the increasing revenue idea.

\subsection{Processes and features}

\begin{figure}[!ht]
		\centering
		\includegraphics[width=1.0\linewidth]{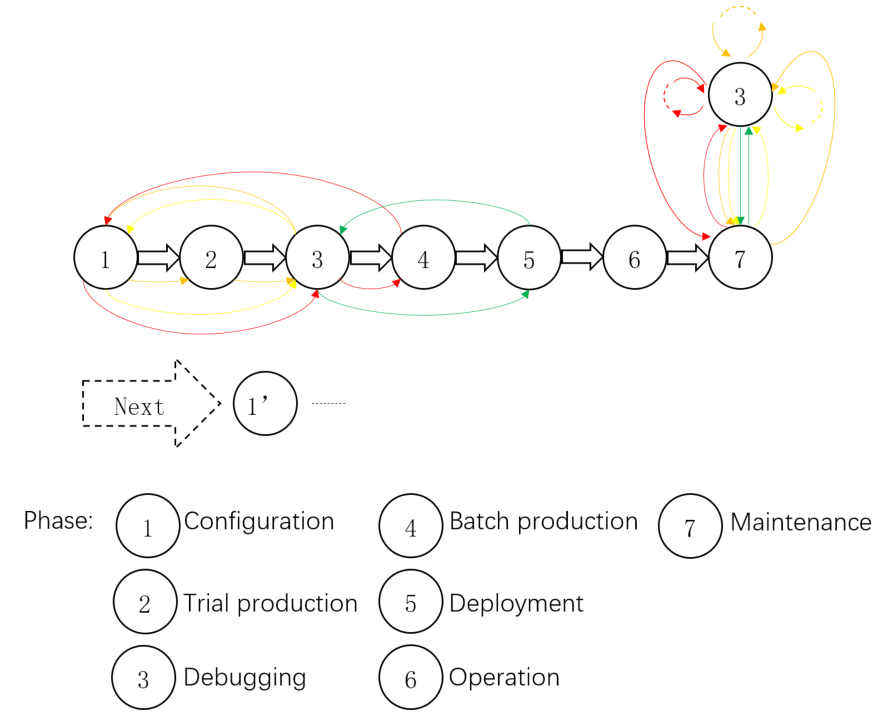}
		\caption[VarPerOptCom]{Life cycle stage flowchart}
		\label{fig:1}
	\end{figure}

Fig. \ref{fig:1} shows a flow chart of the WSN maintainability implementation life cycle stages. The link state body in the life cycle goes as shown by the large arrows. Serial numbers 1 to 7 represent the seven parts of the maintainability implementation life cycle, each link will incur time costs, small arrows for the failure decision processing feedback links, will generate decisions corresponding to the expected resource consumption, the color along the chromatogram from green to red represents the cost from low to high. The study of maintainability is to study ways to reduce costs in the process of changing the state of this WSN project over a ten-year period. The essence is a resource allocation strategy with a high degree of confidence based on certain a priori information (previous engineering experience, technology trends), combined with the state of the project process.

Take the life cycle of a large-scale meter acquisition WSN system as an example, in the configuration stage, design WSN architecture, MAC layer protocol, hardware selection, inter-chip interface, program architecture, etc. And then along the big arrow of the release into the trial production session, according to the previous design of a small number of purchase materials, Circuit board engraving, SMD, test electrical performance, and RF indicators, estimate the yield rate. If the instability has been ruled out as far as possible (the yield rate is reduced in mass production resulting in a large waste of resources and costs), along the big arrow into the debugging stage, according to the design of the architecture to develop the function, according to low-power indicators to test the low-power current. If a problem is found, such as a capacitor indicator cannot reach the data variance distribution of its data sheet so that it cannot meet the electrical performance of WSN products, at this time, through a certain resource cost (set to the second level corresponding to the yellow path according to the estimated scale, a total of four levels) consumption, return to the design session to choose another program or redesign the material selection, and return to the debugging stage to replace the capacitor of the previous finished product material, and conduct the corresponding index test. If there are key components that can not meet the demand so much as to weigh the later cost and redesign debugging production, the latter is lower, such as a material because of force majeure factors can not meet the demand for mass production, then you can take a third level of resource cost expenditure (orange) and then go to the design debugging trial production link. If the trial production stage targets are completed and possible trial and error overheads that are scaled up in the next stage are excluded as much as possible, enter the mass production session. 

We need to be very careful before entering the mass production session, because mass production is an important part of the WSN to amortize costs and maybe a missed session that amplifies errors that bring a lot of invalid expenditures. Therefore, the process expenditure for its failure will be level 4 (red). After successful batch production, enter the deployment link, adjust the performance of WSN to the expected functional indicators (packet loss rate, RSSI, and etc.). In the deployment stage, we install WSN devices on demand in production environments based on pre-existing multi-format, redundant production, so their expenses are often first-tier, so its expenditure is often the first level. And then it is the continuous operation of WSN monitoring and WSN maintenance analysis, troubleshooting, analysis requirements, and upgrade iterations. Switching between maintenance and other stages can be done sequentially according to the failure level and later upgrade requirements and is not limited by the expenditure level. Until the completion of this life cycle after ten years, the lessons learned and reusable resources into the next life cycle.

%
%
%

\subsection{Current situation and challenges}

This subsection summarizes the historical research on maintainability, as well as the work that has been done in the WSN maintainability system, and envisions the challenges that may be faced in future work.

\subsubsection{System}

Maintainability has been extensively studied in the fields of software engineering \cite{mtnblt-sw}, mechanical structures \cite{mtnblt-mech}, and industrial engineering \cite{mtnblt-IE}. 
Maintainability as one of the quality attributes of software engineering, Hadeel Alsolai et al. \cite{mtnblt-sw}  surveyed 56 relevant studies from 35 journals and 21 conferences and showed that the most commonly used software metrics (dependent variables) in the selected primary studies were variable maintenance effort and maintainability index, and most of the studies used class-level product metrics as independent variables. 
Maintainability assessment is the main way to evaluate the maintainability of mechanical product design phase. In order to describe the maintainability of mechanical products, Luo  et al. \cite{mtnblt-mech} proposed a basic assessment object hierarchy and quantitative assessment methods for maintainability design attributes to improve the accuracy and reliability of assessment results. Thus, the accuracy and reliability of the assessment results are improved thus avoiding the weaknesses of the maintainability design. However, this method does not give improvement measures for maintainability weaknesses. 
S. Ahmadi  et al. \cite{mtnblt-IE} divided the performance of the metro construction material transportation system into reliability, maintainability, and availability (RAM) evaluated separately in terms of time as a unit of measurement. The reliability was analyzed to derive the maximum life time of the system and the availability was analyzed and the time of this attribute was the maximum life time of the system for the part of the reliability above the predefined index. To maintain the system above a certain level of reliability, it is necessary to perform To maintain the system above a certain level of reliability, maintainability maintenance needs to be performed, and this process also takes time and is called preventive maintenance interval.

The maintainability evaluation system of WSN is still in the exploration stage as a brand-new idea. Among them, Zhang et al. summarized the top-level design and architecture of maintainability \cite{zqc}. Shen et al. \cite{sj} proposed a preliminary evaluation scheme based on subjective expert experience, hierarchical analysis method, and fuzzy comprehensive evaluation. Gao et al. \cite{gtn} applied an objective sample entropy weight method based on Shen's work to calculate the operational parameters of the project implementation guided by expert opinions in his paper and used it as a basis to update expert experience after the project was completed. Qiu et al. \cite{qct} used confidence rule base for evaluation. Wang \cite{wc} used expert experience combined with EM algorithm and Bayesian network for evaluation. Future resource cost modeling analysis can be performed for the WSN life cycle conducted in parallel.

\subsubsection{Methodology and engineering practice}

The next step in the development of WSN maintainability adopts an alternating iterative approach of theory and experience. The expert experience summarizes the theory, the theory guides the engineering practice, and the experience is summarized from the engineering practice results and then improves the theory.

WSN maintainability from the time dimension to address two issues
: 
\begin{itemize}
\item{How to evaluate the links of this life cycle?}
\item{The data and resources of this life cycle how to guide the next life cycle improvements?}
\end{itemize}
The ultimate goal of solving the problem is to achieve rapid convergence of the corresponding cost to the minimum as the life cycle is rapidly iterated. The way of cost convergence in engineering solutions is still to be explored, and the combination of WSN and Serverless Computing in this paper is an attempt.

The solutions to some problems in maintainability can be based on results from other fields, and in the future can be combined with theories such as optimal control, machine learning, and probability theory to participate in the construction of methodological models. For example, link cost resource consumption and decision process modeling can be attempted in the future using unsupervised learning algorithms to divide attributes and links of factors and explore the contribution of the earlier factors to the later factors at each stage \cite{x-ai}, and then transform the machine learning model into an equivalent neural network with layer-wise-relevance propagation (LRP) for the link parameters in which to explain the contribution of the inter-factor effects.

In terms of engineering practice, there is still a long way to go, but it is possible to make bold assumptions by absorbing ideas from other engineering fields. For example, this paper combines the concept of Serverless Computing and tries to treat WSN as the execution unit of Serverless Computing, and sells it as a distributed system resource by dividing its functions into fine-grained and low-latency operations, thus expanding the revenue space to amortize the cost and increase the revenue.

\section{Serverless Computing}

At this stage, the embedded chip performance has increased significantly and the price has decreased significantly, WSN system design level can be relaxed to consider the performance constraints. In particular, the development of embedded Linux allows the introduction of many excellent design ideas into WSNs, Serverless Computing being one of them. Serverless Computing as, decoupling the development of Serverless Computing application functions from system maintenance abstraction in its WSN domain that is. The following describes the Serverless Computing concepts, WSN maintainability in the use of the features.

\subsection{Concept and application}

Serverless Computing \cite{serverless-computing} is an integration of both BaaS and FaaS as a unified service form to its customers.
 From the perspective of a functional developer it can be seen as Function as a Service (FaaS), and from the perspective of a Serverless Computing provider it can be seen as Backend as a service (BaaS) \cite{gill2022ai}. Serverless Computing allows developers to focus on the development of functional functions without having to focus on the maintenance of the underlying runtime environment. For Serverless Computing vendors, Serverless Computing reduces hardware idle time to improve equipment utilization and achieve greater profitability.
 
 Serverless Computing has been proposed to address the high coupling high cost and inefficiency of the traditional development model lifecycle \cite{XuCSUR2022}. 
 Mondal et al. \cite{sc-application} redefined the Serverless Computing paradigm from an administrator's perspective and integrated it with Kubernetes to accelerate the development of software applications. Theoretical knowledge and experimental evaluations suggest that this new approach can enable developers to design software architectures and development more efficiently by minimizing the cost of maintaining hardware facilities for public cloud providers (e.g., AWS, GCP, Azure). However, serverless functions come with a number of issues such as security threats, cold start problems, and inadequate debugging of functions.
 
 In the WSN space, it can be used as an implementation to improve maintainability and better amortize costs by expanding unit hardware utilization.  
 
\subsection{Serverless Computing used in WSN's maintainability}

The process of Serverless Computing consists of two main aspects, function programming and function services. This idea can guide the development model of WSN maintainability design phase, where users only need to call function services to get the required resources, and the development center of WSN developers shifts from implementing a single function to coordinating the front-end and back-end of the system to implement a function programming environment. The implementation of maintainability may borrow many of its features. Therefore, this subsection summarizes some of the features of Serverless Computing that are used in the WSN maintainability development process and explains how to apply them.

\begin{itemize}
    \item Hostless and elastic: Users do not need to know the specific implementation, operation and maintenance, and troubleshooting of WSN. You can write your own rules to use the acquisition, control, and edge computing resources according to the platform specifications as needed.
    
    \item Lightweight: WSN's hardware resource preferences are highly tied to functionality, facilitating fine-grained segmentation and middleware design. For example, edge computing tasks are done by cortex-A performance cores for statistics, and data collection and device control tasks are collected by cortex-M control cores in accordance with the maximum common multiple of multi-user collection frequency capabilities. This feature also allows users to focus on functional logic when writing requirements. However, due to the development of Serverless Computing at this stage, certain restrictions may be imposed on the user writing specification. Such as the number of functions, the number of loop layers, etc.
    
    \item Short but variable execution times: In the WSN device is reflected as the response of the ad hoc network instructions to the upper user functions, i.e., as the interpretation language of the functions. At the same time, the instructions should be units that can be billed for metrics, and the embedded device can be billed for the number of acquisitions.
    
    \item Burstiness: The task load in Serverless Computing is often difficult to predict and is characterized by burstiness, variable change time, and fluctuating intensity. For the characteristics of WSN distributed, the dynamic design of its topology can be developed in coordination with the front-end serverless service.
    
    \item Migration: The automatic scaling of Serverless Computing platforms often comes with the startup cost of a serviceless function, i.e., the system overhead and service latency can be elevated to the point where inter-system migration becomes unattractive. Resource provisioning at the WSN level has a slightly higher energy overhead on the system due to the low-power requirements of energy-constrained devices, but its lightweight firmware makes migration costs low, and ultimately there is still a tradeoff between the expected benefits of migration and the system resource overhead.
\end{itemize}

\section{Implementation Conception}

For the needs of multiple parameter acquisition and device control in smart cities, we first introduce the common scheme design patterns and problems, and then elaborate on the idea of increasing the usage amortization cost by abstracting the functions achieved by Serverless Computing in our maintainability design.

\subsection{Existing Schemes and Problems}

\begin{figure}[b] 
\sidecaption
\includegraphics[scale=.13]{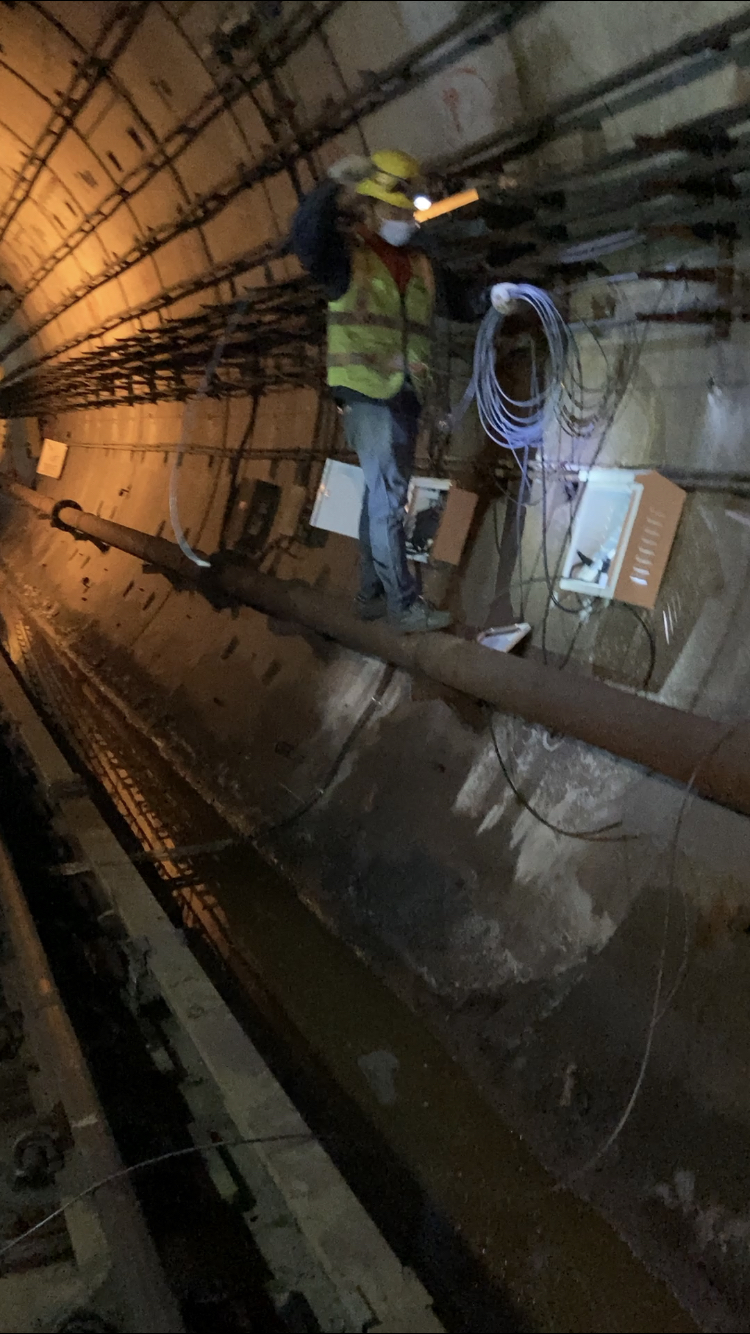}
%
%
\caption{Multiple project departments are responsible for the data acquisition system installation process, which increases the unreliable factors in the industrial environment}
\label{fig:2}       
\end{figure}

According to some commonly used technology of previous experiments, performance, energy, and cost were limited, and developers had to remove codes to reduce functionality and cost expenditures from all aspects. 
Each unit will be its own requirements for separate bids, the common development process of the winning bidder is as follows. In the design phase, they develop WSN for controlling a single device or collecting a single parameter. After debugging and production, at the time of deployment, the WSN products usually occupy a common space with the products of other projects. As shown in Fig. \ref{fig:2}, it is photographed when we evaluated the deployment environment of our WSN product in a tunnel pipe gallery. There are many parameter acquisition equipments deployed in parallel. The overall vertical division of urban wastes public resources, increases system confusion, and leaves security risks.  

\subsubsection{Requirement analysis}

In smart city scenarios, there are many control, acquisition, and edge computing needs. For example, the traffic department's timely response to target detection, dynamic control of tidal lane driving direction, and traffic light timing. Environmental protection department for urban environment monitoring, scientific researchers to obtain the city's sound, light, heat, electromagnetic atmospheric pressure, and other physical parameters and other demand scenarios. 
   We integrate these requirements in the same WSN system, and hardware along with software resources can be configured according to fine-grained demands. To establish a unified project approach to fulfill the development of WSN data acquisition and equipment control function service platform, the system can meet the future increase, modification, and deletion needs. 
Such as the scheme that is not proposed in the tender but may form a profit model in the future with the help of redundant design.

\subsubsection{Hierarchical architecture design}

\begin{figure}[!ht]
		\centering
		\includegraphics[width=0.5\linewidth]{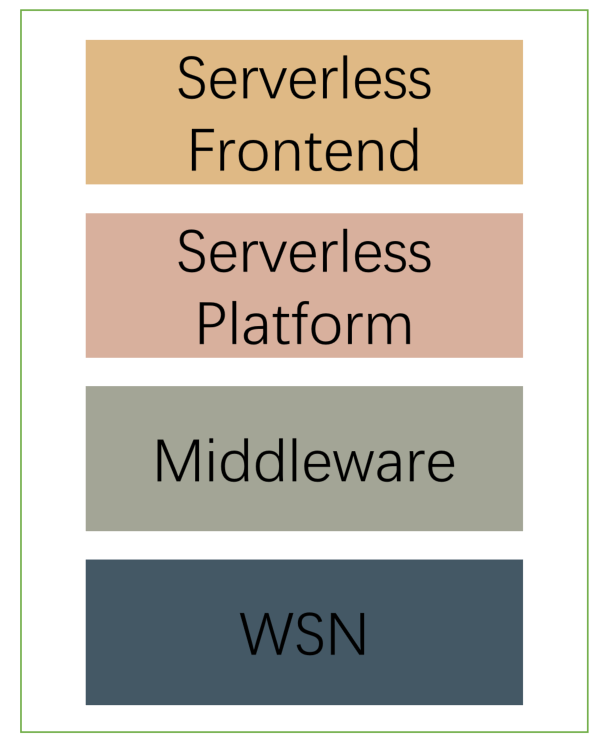}
		\caption[VarPerOptCom]{Layered architecture of Smart city WSN system}
		\label{fig:3}
	\end{figure}

As shown in Fig. \ref{fig:3}, WSN in the resource perspective has the minimum granularity of data collection, device control, edge computing functions. From the perspective of network structure, its minimum granularity for the system's cluster structure can be reorganized on demand. From the hardware selection point of view, its minimum granularity can be achieved by selecting pin-to-pin compatible chips, by selecting resistors and circuit boards to leave redundant material pads. The designed WSN architecture is based on a variant of a star network (consisting of sink gateways, routers, and nodes), considering the platform requirements for function serviceability and the upgrade redundancy to meet the 10-year life expectancy. The middleware acts as a sandbox for the execution of user-developed functions, responsible for both translating the functions into WSN operations (paid resources disassembled into permutations of WSN minimum granularity operations) and returning the corresponding results to the Serverless platform, which is responsible for user management, resource uploads, account billing, user-developed function property management, etc. Serverless frontend is for users to write and upload resource call functions, view resource call results, and record resource usage billing prices.

%
%

\subsubsection{Development process design}

\begin{figure}[b] 

\includegraphics[scale=0.55]{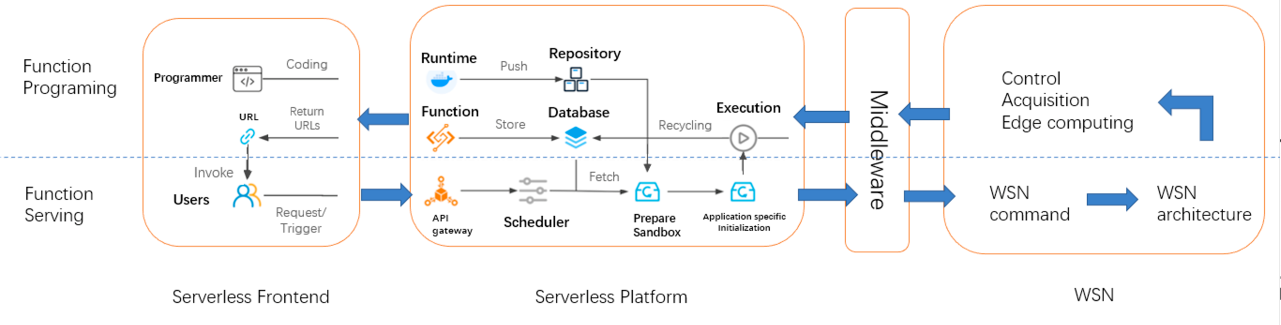}
%
%
\caption{Hierarchical architecture of Smart city system based on Serverless Computing Framework}
\label{fig:4}       
\end{figure}

The overall architectural design details are shown in Fig. \ref{fig:4}, which is derived from \cite{serverless-computing}. The specific implementation details of each part in the life cycle will be described from subsub-section 4.1.4 to 4.1.10.

\subsubsection{System configurations}

\begin{figure}[b] 
\sidecaption
\includegraphics[scale=.4]{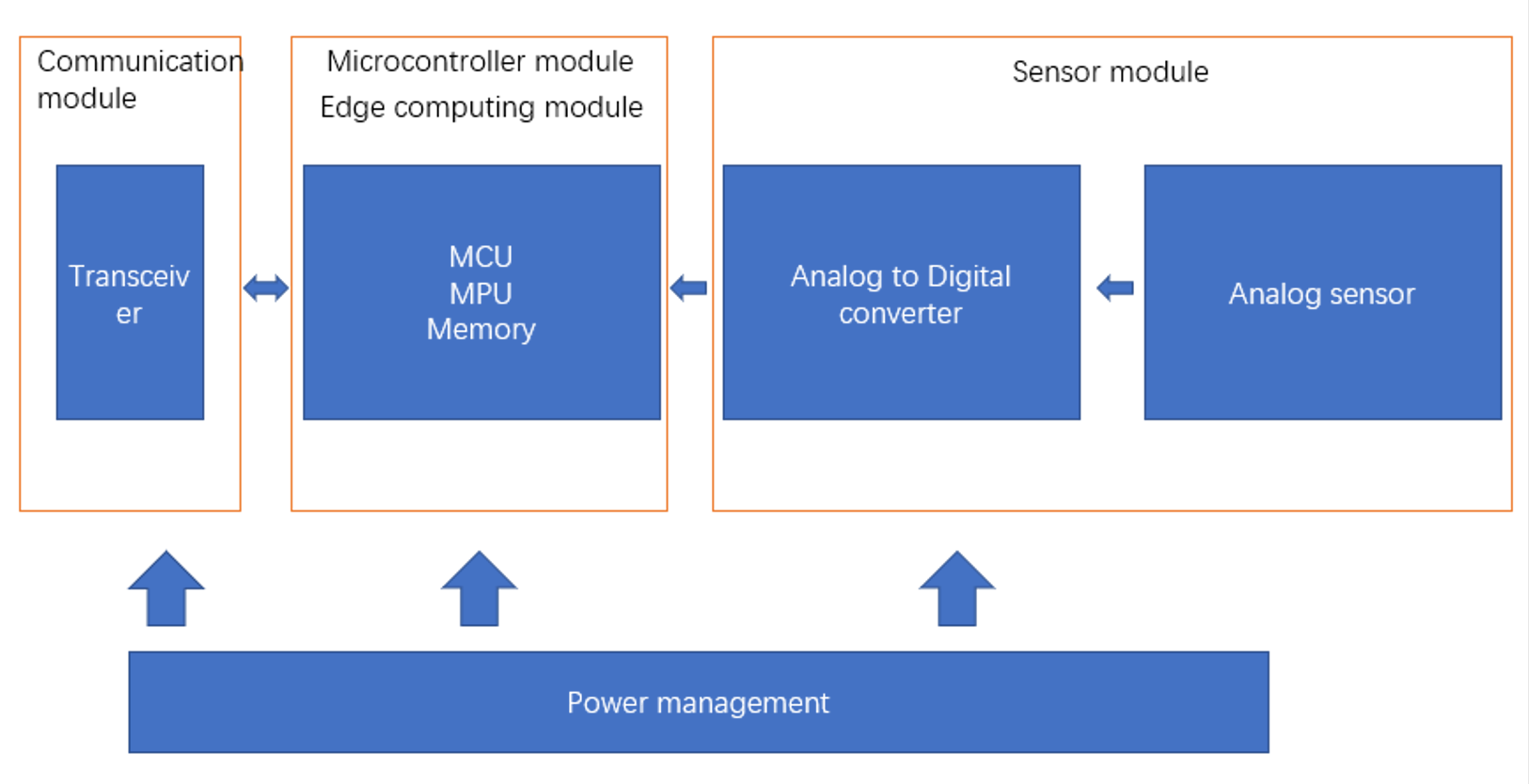}
%
%
\caption{Hardware architecture for WSN}
\label{fig:5}       
\end{figure}

 The modular design can meet the needs of trade-offs between different expected benefits, costs, and functions. The WSN hardware unit is shown in Fig. \ref{fig:5}. The transceiver refers to the RF chip and its supporting peripheral circuits, which serve as the infrastructure of MAC protocol and are responsible for the communication between self-assembled network devices. The micro-control module is divided into single micro control unit (MCU) and micro process unit (MPU), where the sensing module is responsible for acquisition and control functions, and the power management module designs the corresponding circuit according to whether the energy is constrained or not. In constrained devices (e.g. WSN nodes), it is responsible for low-power control. In unconstrained devices (WSN routers and sink gateways), it is responsible for power outage switching backup power, battery side flush and other functions. Feedback speed of hardware resources, communication between hardware can achieve microsecond level completion speed to meet the needs of Serverless microsecond level tasks.
 
 WSN software design focuses on the hardware abstraction layer, driver layer, real-time operating system layer (RTOS), and application layer. The hardware abstraction layer is a circuit-level abstraction of the hardware unit in the software architecture, defined by Yoo S et al. \cite{HAL}, and is intended to solve the error-dead problem when the operating system directly manipulates the chip peripherals, which isolates the system parameter configuration during development from the specific model of peripherals and encapsulates the register behavior of the hardware as a separate function, also known as the board-level support package (BSP). This is to facilitate later development of system functions for specific hardware, i.e., reuse of resources.

The driver layer, as an intermediate link between the operating system layer and the hardware layer, requires efficient and stable code to realize the interactive functions between the upper and lower layers. The operating system layer is often used to make the logic of complex tasks through abstraction and then concise implementation, which also need the efficient execution of program functions. Therefore, the C programming language  has become the least used language in the history of embedded development, and the C programming language  is a mature technology with rich framework code and error handling. But in the scale of ten years, the project often occurs in human resources changes (engineers level inconsistent, engineering details handover incomplete), tight schedule and other problems. Rust, as a result of recent decades of programming language research, has inherent runtime security. It has a comprehensive package management mechanism, and a secure development process in which code that is not securely specified cannot be checked by the compiler. And it follows a zero abstraction overhead enabling developers to securely control the underlying capabilities. Therefore, we boldly suggest that you can try Rust\&C hybrid development in both layers.

The open source, widespread use of Linux systems has enabled the rapid migration and deployment of existing cutting-edge engineering technologies. The increased resource density of microcontroller chips has also enabled complex operating systems like Linux to be applied in embedded devices. Therefore, in our design, arithmetic, energy-unconstrained devices are implemented using a mixture of Linux system and RTOS, the former is responsible for arithmetic, business logic, and the latter is responsible for WSN protocol stack, sensor acquisition, and peripheral control. Resource-constrained devices, energy-unconstrained devices use RTOS to implement the sink-gateway routing node function of WSN, and energy-unrestricted devices also use RTOS to implement the node and sensor acquisition function of WSN.

WSN application layer running Serverless Computing middleware backend, i.e., WSN devices with fine-grained control commands, including but not limited to self-organizing network topology adjustment, control, acquisition, and edge computing resource invocation. Users can subscribe on-demand and use on-demand in accordance with the function development.

\subsubsection{Middleware configurations}

The middleware runs as a service on the service cluster to realize the conversion of user functions and WSN instructions and the uploading of WSN execution results. Instruction conversion means functions that are involved by multiple users, disassembled into efficient execution of WSN instructions, upload refers to the execution results in accordance with the user function resource calls to split and upload. For example, two user subjects need WSN collection resources to collect air pollutant density, the former according to the frequency of 5 minutes to 100 nodes of data collection once, the latter according to 10 minutes to collect 100 nodes, but need to be above the threshold to trigger an alarm and upload data. After aggregating the two functions that need to be executed, the middleware will merge the commands for their execution and send the instructions to WSN for 100 nodes running at a frequency of once every 5 minutes and set the threshold value to trigger sending, which is the minimum resource consumption for executing all users' commands. The middleware will then upload the results on demand according to its needs, and the Serverless Computing platform will be billed according to the resources invoked.

\subsubsection{Serverless platform configuration}
 
Serviceless computing is divided into platform and frontend, where the platform side runs the user identification and platform dynamic scaling services for Serverless Computing, and the frontend runs the services used to write functions and visualize results. From a functional programming point of view, the user (programmer) writes functions (cycle execution, event triggering, conditional judgments) in the local environment provided by the frontend (possibly a page in a browser for WSN users) according to specifications we define, and the functions are submitted to the platform's function database, which then returns an interface to call the functions The function is then submitted to the platform's function database, which returns an interface to call the function. From the function service perspective, when a user triggers or invokes a function through the interface, the platform obtains the function from the function database and performs code compliance and security checks, and then sends its instructions to the middleware platform, which aggregates function functions from multiple users and then merges the requirements into one category, breaking them down into WSN instructions in the most 
efficient way possible.

Considering the design scheme as above, according to the demand analysis of subsection 4.2, we plan to use 10 energy-unrestricted edge computing WSN hardware responsible for target detection, tidal lane traffic light time dynamic control condition judgment, 40 energy-unrestricted devices (gateway-sink nodes-routing) responsible for, 100 energy-restricted low-power devices to achieve coverage of a small-scale city's major traffic Serverless Computing platform and middleware are deployed as software services on a mature cloud computing platform or on self-built servers.

\subsubsection{Trial production}

The pilot production phase follows multiple design options to create separate hardware, purchase servers, and develop a Serverless Computing platform. The purpose is to verify whether the design is feasible, and to stop it in time if it is not. The trial production should fully consider whether the required materials are available and stable in the current market environment, and whether the performance meets its datasheet nominal specifications. It is also necessary to evaluate the tradeoff between the cost and benefit of alternative models, the tradeoff between redundant designs and the benefit of future WSN implementation functions, and the tradeoff between equipment life and maintenance over a ten-year life cycle. Platform development considers the tradeoff between the cost of running the platform and performance redundancy, and the tradeoff between the cost of dynamic platform deflation control and the efficiency of system operation. For example, whether the designed post-paralysis instructions reach the balance interval after the trade-off of resources, cost, and stability. Whether the designed WSN key components meet the data book metrics, i.e., whether they meet the balance between expected life and maintenance costs. In case of unanticipated conditions, the design is adjusted until there are no significant problems. For the needs of smart cities, we produce a prototype of the WSN hardware in the design phase separately by type, and the Serverless Computing platform is developed according to the designed architecture.

\subsubsection{Debugging}

After completing the trial production, the software and hardware are tested for communication and compatibility respectively. Server-side debugging function development running platform. Independent debugging and joint debugging are selected in turn according to the development progress. In the early stages of development, hardware and software independent debugging, WSN side debugging hardware performance until approximating the datasheet indicators. WSN software based on the QEMU virtual machine simulation of the kernel, peripherals, interrupt development debugging control core, and RTOS functions. The stability of the Serverless Computing platform, security, and the ability to complete the complete isolation of the user from the details of the WSN device. whether the Serverless Computing server meets the rated amount of concurrent user requests. Whether the middleware correctly subsumes and parses user functions, and whether the results of resource calls are correctly uploaded on demand. 
WSN hardware performance to meet the data manual indicators, WSN software program operating logic is normal and through the sample test, Serverless Computing platform and middleware of the above indicators are verified and function properly, and all parts of the trouble-free operation time respectively to meet their respective design requirements. 
After all the respective debugging is completed, the joint debugging of hardware and software can be performed later. With the experience of foregoing practice, WSN devices produced at 30\% of the batch production can meet the basic experimental scenario commissioning requirements. The purpose is still to reduce the errors that may be amplified in the mass production phase.

\subsubsection{Batch production}

Verify the large-scale communication function to ensure that the communication success rate is above a certain standard, and test the platform function of service-free computing after its verification is successful. Select the program with expected benefits and cost trade-offs to meet the predetermined targets for mass production. And after mass production, all tests should be conducted to determine whether there are failures beyond those expected in the trial production and debugging phases. For example, whether the statistics on RF performance of WSN nodes after mass production are in line with the data book distribution, whether the middleware and WSN protocols can meet the predefined execution efficiency when dynamically expanded, and whether there are problems with the Serverless Computing platform for concurrency of the actual scale.

\subsubsection{Deployment, operation, and maintenance}

The local communication environment in the field deployment phase may have the greatest impact on the system. Therefore, the deployment environment communication channel needs to be measured before deployment, and the WSN switches to the solution with the highest communication success rate, and the server-side Serverless Computing platform comes online to provide services to customers. The operational phase may provide some information to us, which provides a priori information for the selection of the maintenance strategy for the current cycle and experience for the design of the next life cycle iteration. Therefore the fault phenomenon and cause are crucial. 
So we use a sniffer system here. The maintenance phase also follows the cost reduction and efficiency increase strategy. The sniffer device \cite{sniffer} continuously monitors and analyzes the WSN operational status to improve the WSN protocol stack. Node failure requires maintenance priority to use Sniffer to send a restart command to WSN, if not work then choose the more expensive field troubleshooting to replace the faulty equipment. This is cycled between links until the end of the 10-year life cycle.

\subsection{Resource Provisioning strategy}

\subsubsection{Initial time and time period}

The maximum frequency of equipment acquisition without resource constraint first as the most fine-grained acquisition resource division unit of W. The period is divided into pieces, and the resource-constrained equipment performs as little as possible. For example, assume that the node acquires at most 600 times a minute. Then it can be a time anchor point as a benchmark to 600 users to sell once a minute or 300 users to sell half a minute acquisition task.

\subsubsection{Task preemption priority and success rate}

Interruption priority is assigned according to the importance of the task. The high priority is expensive and guarantees a success rate of more than 99\%. For example, tidal lane control uses the highest interruption priority highest rate, the rest of the collection is sensitive to the time electric cycle in the second priority second rate (the collection process allows a certain failure rate), and finally the data volume requirements, time is not sensitive to the task of scientific research collection to take the lowest priority interruption lowest rate (in the hardware resource occupation low when the collection, collection time, cycle of the lowest guarantee).

\subsubsection{Profit from prepaid costs}

Functional resource calls create the possibility for continued profitability at a later stage. After we implement the basic function execution framework, this approach allows users to fully explore the most efficient way to achieve their needs and guides us to analyze the needs from the user's perspective. At the same time, the shared ownership of intellectual property between the user and us makes the user willing to put effort into iterating and upgrading the model to enrich the system, and we are able to sell functionality to other customers, thus increasing the resources available in the next lifecycle and expanding our profitability.

\section{Conclusions and Future Directions}

First, the maintainability of WSN and serviceless computing are introduced. The implementation of serviceless computing in WSN maintainability is introduced for the problem of optimal lifecycle cost and resource consumption of WSN to achieve maximum profitability. Finally, based on the scenario of a smart city gives the whole process solution in the life cycle, firstly, the design phase integrates the requirements of customers from all parties in the same WSN system as far as possible, isolates the system details from the users, and only provides the resource invocation interface that conforms to the specification. Trial production, debugging, mass production phase to improve the hardware and software implementation details. Deployment, operation, maintenance phase in the ten-year life of the premise of the link switch decision and continuous summary of experience, the production of reusable resources.

In the concept of Serverless Computing, hardware systems that meet life expectancy, high reliability, and performance redundancy are designed through the concept of maintainability from the perspective of the infrastructure provider. From the service provider's point of view, the functionality is used in a user-programmable way by reusing hardware resources and paying for them on demand. From the user's point of view, shielding the back-end implementation details only requires the uploading of resource invocation patterns written according to specifications in the form of a web server or sandbox according to their needs. At the same time the user-created and fine-grained WSN operations as well as intellectual property rights and updates. It can reduce the coupling in the production environment, reduce resource waste and improve system security. 

As future research direction, although the concept and implementation plan proposed in this paper are innovative, it relies on the development of the existing technology, and the biggest controversy may be in today's industry cost distribution, profit model. However, with the development of embedded hardware and software technology and cost reduction, in the current Moore's Law development, the tendency is also from technical progress to system improvement, that is, to explore higher resource utilization efficiency within the same system, such as pipelining technology of instruction set, virtualization technology of operating system, shadow page table in virtual machine, etc., essentially to improve the resource utilization in each level. In order to meet the needs of various industries and fields, we believe that cost reduction, architecture optimization, and quality improvement will become the future research direction in engineering technology if the cost and technology development is suitable for the concept of this paper.

\end{document}